\documentstyle{article}
\begin{document}
\title{Spin and Rotation in General Relativity} 
\author{Lewis H. Ryder\\
Physics Laboratory, University of Kent at Canterbury\\
Canterbury, Kent CT2 7NR, UK\\ E-mail:  l.h.ryder@ukc.ac.uk\\
\and 
Bahram Mashhoon\\
Department of Physics and Astronomy,\\ University of
Missouri-Columbia\\ Columbia, Missouri 65211, USA\\ E-mail: 
mashhoonb@missouri.edu}
\maketitle

\section{General remarks}

Broadly speaking, the ``role of spin and rotation in General Relativity" covers two topics;  the behavior of spinning particles in GR --- this ``spin" being
either classical or quantum mechanical, and the physics associated with ({\it noninertial}) rotations.  The papers presented to this session cover  both
these aspects of the subject.  To the non-specialist, the most familiar heading in this general area is the Lense-Thirring effect, a precessional effect
which is predicted (though so far not observed) to take place close to a rotating body. The papers of Ciufolini and Teyssandier are both devoted to this
effect.  Also well-known is the (gravitomagnetic) clock effect; as its name implies, this is concerned with time nonintegrability, rather than with
precession.  Tartaglia and Maleki both address this topic; and we live, moreover, at a time when both the Lense-Thirring and gravitomagnetic clock effects
have the enticing possibility of experimental confirmation in the near future.

Connections with gravitational waves (also soon to be detected?) are
made, though in very different ways by Cooperstock, Slagter, and Suzuki and Maeda.  The
behavior of spinning particles in various types of gravitational fields is analyzed by  Mohseni, Tucker and Wang, by Mukhopadhyay and by Singh and Papini,
the latter two contributions being concerned specifically with Dirac particles, while White and Raine --- alone in this session --- consider a theory with
torsion.  Bini, Gemelli and Ruffini give a more general account of the description of spinning particles in GR.  In a nonquantum context, Wu and Xu
include rotational motions in their analysis of hydrodynamic equations in General Relativity.  The second subject of this session, concerned with rotation
effects, has spawned, in recent years, the interesting phenomenon of spin-rotation coupling.  This effect is described in the following sections of this
introduction, and provides the motivation for Ryder's paper, which shows that it is consistent with special relativity. 

\section{Spin-rotation coupling}

In classical mechanics, the inertial properties of matter can be illustrated through numerous phenomena involving frames of reference undergoing
translational and rotational accelerations.  In nonrelativistic quantum mechanics, the state of a particle is primarily characterized by its inertial
mass.  On the other hand, the irreducible unitary representations of the inhomogeneous Lorentz group are characterized by both mass and spin.  Thus in
relativistic quantum mechanics two independent types of inertia are expected : the inertia that is primarily due to mass-energy and the purely quantum
inertia due to intrinsic spin.  The phenomenon of spin-rotation coupling illustrates the inertia of intrinsic spin.  

Let us first consider a real massless scalar field $\Phi(x)$ in Minkowski spacetime such that $\Box\Phi = 0$.  Imagine an observer rotating with frequency
$\Omega$ about the $z$-axis in the $(x, y)$-plane.  The coordinate transformation to the rest frame of the rotating observer is given by
$t^{\prime}=t, z^{\prime} = z$, and 
\begin{eqnarray} x^{\prime} & = & x\; {\rm cos}\;\Omega t + y\; {\rm sin}\;\Omega t\;\;,\cr y^{\prime} & = & -x\; {\rm sin}\;\Omega t + y\; {\rm
cos}\;\Omega t\;\;.
\end{eqnarray}
We note that this coordinate transformation in fact coincides with the geodesic frame set up along the worldline of the noninertial observer that
remains at rest at the origin of spatial coordinates but refers observations to the spatial axes rotating with frequency $\Omega$.  In terms of the
spherical polar coordinates $(r,\theta, \varphi)$, this transformation can be expressed as $r^{\prime} = r,
\theta^{\prime} = \theta$ and $\varphi^{\prime} =
\varphi - \Omega t$.  We suppose that the scalar field can be expressed in terms of a superposition of Fourier modes as
\begin{equation}
\Phi(x) = {\rm Re}\sum_{\bf k}\;\alpha({\bf k})\;e^{i({\bf k}\cdot{\bf r} - \omega t)}\;\;,
\end{equation}
where $\omega = ck$.  Writing a plane wave in terms of spherical waves
\begin{equation} e^{i{\bf k}\cdot{\bf r}} = 4\pi\sum_{lm}\;i^l\:j_l(kr)\;Y^{\ast}_{lm}({\bf \hat k})\;Y_{lm}({\bf \hat r})\;\;,
\end{equation}
we find that upon transformation to the rotating system of the observer,
$\Phi(x) = \Phi^{\prime}(x^{\prime})$ and that 
\begin{equation}
\Phi^{\prime}(x^{\prime})=4\pi{\rm Re}\sum_{{\bf k}lm}\;\alpha({\bf k})i^lj_l(kr)\;Y^{\ast}_{lm}({\bf\hat k})\;Y_{lm}({\bf\hat
r}^{\prime})e^{-i(\omega-m\Omega)t}\;\;.
\end{equation}
Here $j_l, l = 0, 1, 2, ...$, are spherical Bessel functions and we have used the fact that for spherical harmonics 
\begin{equation} Y_{lm}(\theta, \varphi) = e^{im\Omega t}\;Y_{lm}(\theta^{\prime},
\varphi^{\prime})\;\;.
\end{equation}
Thus the frequencies of the modes as measured by the rotating observer are given by $\omega^{\prime} = \gamma(\omega - m\Omega)$, where $m = 0,
\pm1, \pm 2, ...$, and the Lorentz factor $\gamma$ accounts for time dilation since $dt = \gamma d\tau$.  Comparing this result with the Doppler effect
$\omega^{\prime}_{D} = \gamma(\omega - {\bf v}\cdot{\bf k})$, we find from ${\bf v} = {\bf \Omega}\times {\bf r}$ that
$\omega_D^{\prime} = \gamma(\omega - {\bf l}\cdot{\bf \Omega})$, where ${\bf l} = {\bf r}\times{\bf k}$ and $\hbar {\bf l} = {\bf L}$ is the (orbital)
angular momentum.  That is, our result for
$\omega^{\prime}$ goes over to $\omega^{\prime}_{D}$ in the JWKB limit.

Let us next consider an electromagnetic field in the inertial frame characterized by the vector potential in a gauge such that $A_{\mu} = (0, {\bf A})$ with 
\begin{equation} {\bf A}(x) = {\rm Re}\sum_{\bf k}\;{\bf a}({\bf k})e^{i({\bf k}\cdot{\bf r} - \omega t)}\;\;,
\end{equation}
where $\omega = ck$ as before and ${\bf k}\cdot{\bf a}({\bf k}) = 0$ by transversality.  The transformation of the vector potential to the
rotating frame implies that $A^{\prime^0}(x^{\prime})=0$ and ${\bf A}^{\prime}(x^{\prime}) = {\bf A}(x)$.  Writing ${\bf A}(x)$ in terms of vector spherical
harmonics, we find
\begin{equation} {\bf A}(x) = 4\pi\:{\rm Re}\;\sum_{{\bf k}JlM}i^{l}\:\left[{\bf Y}^{\ast}_{JlM}({\bf\hat k})\cdot {\bf a}({\bf k})\right]j_l(kr){\bf
Y}_{JlM}({\bf\hat r})e^{-i\omega t}\;\;,
\end{equation}
where $J$ and $M$ are the total angular momentum parameters of the field with
$J = 1, 2, 3, ...$, and $M = 0, \pm 1, \pm 2, ...\;.$  It follows that in the rotating system we have 
\begin{equation} {\bf A}^{\prime}(x^{\prime}) = 4\pi\:{\rm Re}\sum_{{\bf k}JlM}i^l\:\left[{\bf Y}^{\ast}_{JlM}({\bf\hat k})\cdot{\bf a}({\bf
k})\right]j_l(kr){\bf Y}_{JlM}({\bf\hat r}^{\prime})e^{-i(\omega-M\Omega)t}\;\;,
\end{equation}
so that as before the new frequencies are
$\omega^{\prime}=\gamma(\omega - M\Omega)$ with $M = 0, \pm 1, \pm 2, ...$, for each mode with frequency $\omega$ in the inertial frame. 

To distinguish the role of the intrinsic spin of the photon from that of the orbital part in the formula for $\omega^{\prime}$, let us consider the case of
normal incidence, where the orbital contribution vanishes.  To this end, we consider a circularly-polarized plane monochromatic electromagnetic wave
propagating along the
$z$-axis given by
\begin{equation} {\bf A} = {\rm Re}\left[a_0({\bf\hat x}\pm i{\bf\hat y})e^{ik(z-ct)}\right]\;\;.
\end{equation}
Here the upper sign represents positive helicity (RCP) radiation, while the lower sign represents negative helicity (LCP) radiation.  Since ${\bf
A}^{\prime}(x^{\prime}) = {\bf A}(x)$, we have
\begin{eqnarray} A^{\prime}_1 &=& A_1\;{\rm cos}\;\Omega t + A_2\;{\rm sin}\;\Omega t\;\;,\cr A^{\prime}_2 &=& -A_1\;{\rm sin}\;\Omega t + A_2\;{\rm
cos}\;\Omega t\;\;,\cr A^{\prime}_3 &=& A_3\;\;,
\end{eqnarray}
so that according to the rotating observer the coordinate components of the vector potential are given by
\begin{equation} {\bf A}^{\prime} = {\rm Re}\left[a_0({\bf\hat x}^{\prime}\pm i{\bf\hat y}^{\prime})e^{ikz^{\prime}-i(\omega\mp\Omega)t^{\prime}}\right]\;\;.
\end{equation}

To determine the physically measurable field in the rotating frame, we must project the components of the field $F^{\prime}_{\mu\nu} = A^{\prime}_{\nu,
\mu}-A^{\prime}_{\mu,
\nu}$ on the orthonormal tetrad frame of the observer.  Let us consider a rotating observer following the path
\begin{equation} x = \rho\;{\rm cos}\;(\varphi_0 + \Omega t)\;\;,\;\;y=\rho\;{\rm sin}\;(\varphi_0 + \Omega t)\;\;,\;\;z = z_0\;\;.
\end{equation}
The tetrad frame of the rotating observer with respect to the inertial frame is given in $(ct, x, y, z)$ coordinates by
\begin{eqnarray}
\lambda^{\mu}_{(0)} &=& \gamma(1,\;\; -\beta\;{\rm sin}\;\varphi\;\;,\;\;\beta\;{\rm cos}\;\varphi\;\;,\;\;0)\;\;,\cr
\lambda^{\mu}_{(1)} &=& (0,\;\;{\rm cos}\;\varphi\;\;,\;\;{\rm sin}\;\varphi\;\;,\;\;0)\;\;,\cr
\lambda^{\mu}_{(2)} &=& \gamma(\beta,\;\;{\rm -sin}\;\varphi\;\;,\;\;{\rm cos}\;\varphi\;\;,\;\;0)\;\;,\cr
\lambda^{\mu}_{(3)} &=& (0,\; 0,\; 0,\; 1)\;\;,
\end{eqnarray}
where $\beta = \rho\Omega/c\;\;,\;\; \gamma = (1-\beta^2)^{-1/2}$ and
$\varphi = \varphi_0+\Omega t$.

In the rotating frame, the observer has fixed spatial coordinates
\begin{equation} x^{\prime} = \rho\:{\rm cos}\:\varphi_0\;\;,\;\;y^{\prime} = \rho\:{\rm sin}\:\varphi_0\;\;,\;\;z^{\prime} = z_0\;\;,
\end{equation}
and the corresponding tetrad frame in $(ct^{\prime}, x^{\prime}, y^{\prime}, z^{\prime})$ coordinates is independent of time
\begin{eqnarray}
\lambda^{\prime\mu}_{(0)} &=&\gamma(1,\; 0,\; 0,\; 0)\;\;,\cr
\lambda^{\prime\mu}_{(1)} &=& (0,\;\; \frac{x^{\prime}}{\rho}\;\;,\;\;
\frac{y^{\prime}}{\rho}\;\;,\;\; 0)\;\;,\cr
\lambda^{\prime\mu}_{(2)} &=& (\gamma\beta,\;\;-\frac{y^{\prime}}{\rho\gamma}\;\;,\;\;\frac{x^{\prime}}{\rho\gamma}\;\;,\;\;0)\cr 
\lambda^{\prime\mu}_{(3)} &=& (0,\; 0,\; 0,\; 1)\;\;,
\end{eqnarray}
where $\rho = (x^{\prime 2}+y^{\prime 2})^{1/2}$.  Let us note that the metric in the rotating frame is given by
$-ds^2=g^{\prime}_{\mu\nu}\:dx^{\prime \mu}dx^{\prime \nu}$, where $(c=1)$
\begin{equation}  (g^{\prime}_{\mu\nu}) = \left[ \begin{array}{cccc} -1+\Omega^2(x^{\prime 2}+y^{\prime 2}) & -\Omega y^{\prime}& \Omega x^{\prime} & 0\\ 
-\Omega y^{\prime} & 1 & 0 & 0\\ 
\Omega x^{\prime} & 0 & 1 & 0\\ 0 & 0 & 0 & 1
\end{array} \right]\;\;.
\end{equation}
The rotating coordinates are admissible within the light cylinder at
$r = 1/\Omega$.  

It is straightforward to work out the projection of the field, derived from (8), on the tetrad frame of the observer (15); alternatively, this calculation
could be done in the inertial frame resulting in
$F_{(\alpha)(\beta)}=F_{\mu\nu}\lambda^{\mu}_{(\alpha)}\lambda^{\nu}_{(\beta)}$.  Either way, the Fourier analysis of the measured field in the {\it
uniformly} rotating frame,
$F_{(\alpha)(\beta)}$, implies that the measured frequency spectrum of the radiation is given by $\omega^{\prime} =
\gamma(\omega - M\Omega)$, where $M = 0, \pm 1, \pm 2, ...\;.$  In particular, for the special case of radiation propagating along the axis of rotation
(11), we find
$\omega^{\prime} = \gamma(\omega\mp\Omega)$.  This has a simple physical interpretation:  the electric and magnetic fields rotate in the positive (negative)
sense with frequency $\omega$ along the direction of propagation of an RCP$\;$(LCP) wave; therefore, the rotating observer will perceive the radiation
frequency to be
$\omega -
\Omega\;(\omega+\Omega)$ with respect to coordinate time in accordance with equation (11).  This result should be compared and contrasted with the
transverse Doppler effect $\omega^{\prime}_{D} = \gamma\omega$ : in the JWKB limit $\Omega/\omega =
\lambda/(2\pi c/\Omega)\rightarrow 0$, the influence of the intrinsic spin can be neglected, and we recover the transverse Doppler formula of the standard
theory of relativity [1,2].

The Dirac equation for a spin $1\over 2$ particle will be treated in Section 4, but the general result is basically the same, namely,
$\omega^{\prime}=\gamma(\omega - M\Omega)$, where $M\mp 1/2 = 0, \pm1, \pm 2, ...\;.$

Let us now consider the physical consequences of the spin-rotation coupling.  We note that our general result is {\it inconsistent} with {\it phase
invariance} that leads to the Doppler and aberration effects.  That is, only in the JWKB limit is the phase invariance of the wave recovered.  As a
consequence of this deviation from phase invariance, it is possible for an observer to rotate in such a way that
$\omega^{\prime}=0$, i.e. for $\Omega = \omega/M$ the wave stands completely still.  By a mere rotation, an observer can stay at rest with respect to a
fundamental radiation field.  This is a basic consequence of the hypothesis of locality that underlies the theoretical determination of what accelerated
observers measure in the standard interpretation of the theory of relativity.  Let us recall that
$\omega^{\prime} = 0$ is excluded in the Doppler effect since $v<c$; however, the Doppler effect is in general valid only in the JWKB limit.  A nonlocal
theory of accelerated observers has been developed with the intention of preventing accelerated observers from comoving with basic radiation fields [3]. 

What about the possibility that $\omega^{\prime}$ can be negative?  The issue that we need to discuss here is whether the rotating observer could interpret
such modes as positive frequency radiation propagating in the opposite direction.  To see what such an interpretation would signify, let us imagine a
quantum system that undergoes a transition from a stationary state characterized by $(E_1, J_1, \hbar M_1)$ according to standard inertial observers to a
lower energy state $(E_2, J_2, \hbar M_2)$ by emitting a photon of energy $E_1 - E_2 = \hbar\omega$.  The photon is detected by the rotating observer and is
found to have energy $\hbar\omega^{\prime} =
\gamma(\hbar\omega-\hbar M\Omega)$.  That is, from the viewpoint of the rotating observer the states have energies $E^{\prime}_{i} = \gamma(E_i - \hbar
M_i\Omega), i = 1, 2,$ and $E^{\prime}_1 - E^{\prime}_2 = \hbar\omega^{\prime}$ with $M_1 - M_2 = M$.  If $\omega^{\prime} < 0$, the rotating observer could
claim that the photon was emitted by its detector and was absorbed by the state with energy
$E^{\prime}_1$  causing a transition to the state of higher energy
$E_2^{\prime}$.  The only difficulty with this approach is that the causal sequence of events from the emission of the photon to its absorption is reversed
whenever
$\omega^{\prime}<0$.  To preserve the causal sequence of events according to all observers, one would have to admit the possibility of negative energy
states according to accelerated observers.  No basic difficulty is encountered in this way, since the relativity of motion does not extend to accelerated
systems.  In relativity theory acceleration can be characterized in an absolute, i.e. coordinate-invariant and observer-independent, manner.  For instance,
if $a^{\mu} = Du^{\mu}/d\tau$ is the acceleration of an observer, then $a^{\mu}a_{\mu} = -g^2$, where $g>0$ is the magnitude of the acceleration.  Once
$g=0$, then $a^{\mu}=0$ and the observer is not accelerated; otherwise, the observer has {\it translational} acceleration
$g(\tau)$.

For a rotating observer, the rotational coupling of the total angular momentum naturally splits into the orbital and the intrinsic spin parts.  In the case
of electromagnetic radiation, the orbital coupling has been known in its limiting Doppler (i.e. JWKB) form for almost a century and is commonly referred to
as the ``Sagnac effect'' [4].  On the other hand, the observational consequences of the intrinsic spin-rotation coupling have been explored for the past
several decades [5].

\section{Observational aspects of the spin-rotation coupling}

In most experimental circumstances the rotation frequency of the reference frame
$\Omega$ is much smaller than the (de Broglie) frequency of the particle involved; therefore, $\Omega/\omega\ll 1$ in the general formula
$\omega^{\prime}=\gamma(\omega - M\Omega)$ discussed before.  Hence it is useful to consider the JWKB approximation for the purpose of discussing
experiments.  To this end, we can write the energy measured in the rotating frame as 
\begin{equation} 
E^{\prime}_{\bf v} = \gamma(E - {\bf L}\cdot{\bf \Omega}) - \gamma\; {\bf S}\cdot{\bf
\Omega}\;\;.
\end{equation}
It is possible to write a corresponding formula for the momentum of the wave as measured in the rotating frame [6]
\begin{equation} {\bf P}^{\prime}_{\bf v} = {\bf P} + (\gamma - 1)({\bf
P}\cdot{\hat{\mbox{\boldmath$\beta$}}})\;{\hat{\mbox{\boldmath$\beta$}}} - {1\over c}
\gamma E {{\mbox{\boldmath$\beta$}}} + {1\over c}\gamma({\bf S}\cdot{\bf{\Omega}}){{\mbox{\boldmath{$\beta$}}}}\;\;.
\end{equation}
To see where the spin-rotation coupling term in (18) comes from, imagine the noninertial observer at rest at
the origin of the rotating coordinate system; in this case, $E^{\prime}_{0}=E-{\bf S}\cdot{\bf{\Omega}}$ and ${\bf P}^{\prime}_{0}={\bf P}$.  In the eikonal
approximation, any other observer at rest in the rotating frame is expected to be different from the one on the rotation axis by only a boost with
${\bf v}=c{{\mbox{\boldmath{$\beta$}}}} = {\bf{\Omega}}\times{\bf r}$.  In most observational situations $\beta\ll 1$, so that the nonrelativistic
approximation is adequate for the motion of the observer.  Moreover, throughout this section we assume that ${\bf{\Omega}}$ is constant in time.  

The pointwise determination of the energy and momentum of the particle in (17) and (18) illustrates the assumption of locality in the theory of
relativity [2].  To describe physics in a rotating frame of reference, however, we need to incorporate these local ``pointlike'' frames of reference into a
consistent extended reference frame that would be appropriate for the description of experiments.  To this end, we choose the geodesic reference frame
given in equation (1) that is established along the worldline of the noninertial observer at rest on the axis of rotation and note that in this reference
frame ${\bf P}^{\prime} = {\bf P}$ in the sense of equations (10) and $E^{\prime} = E - {\bf L}\cdot{\bf{\Omega}} - {\bf
S}\cdot{\bf{\Omega}}$. The energy and momentum of the particle in the rotating frame determine --- in our approximation scheme --- the phase
$\phi$ of the wave function representing the state of the particle in the rotating frame, i.e.
\begin{equation}
\hbar\; d\phi = -E^{\prime}\;dt^{\prime} + {\bf P}^{\prime}\cdot d{\bf x}^{\prime}\;\;.
\end{equation}
The variation of the phase can be measured via interferometry, which involves the detection of the interference of two or more coherent waves.  Let
$\phi_{D}=\phi(t^{\prime}_{D}, {\bf x}^{\prime}_{D})$ be the phase of the wave at the detector in an interferometer, then 
\begin{equation}
\phi_D = \phi_S + \int^D_S\:(-E^{\prime}dt^{\prime} + {\bf P}^{\prime}\cdot d{\bf x}^{\prime})\;\;,
\end{equation}
where $\phi_S = \phi(t^{\prime}_S, {\bf x}^{\prime}_S)$ is the phase of the wave at the source.  Imagine a beam of radiation that is {\it
coherently} split at the source into two beams that are then brought back together to interfere such that a phase shift $\Delta\phi$ is detected, where
$\Delta\phi = \phi^{+}_D - \phi^{-}_D$ is the phase difference between the two beams at the detector.  Here $\phi^{+}(\phi^{-})$ refers to the phase along
the counterclockwise (clockwise) beam.

The coupling of the orbital angular momentum of the particle with the rotation of the reference frame is revealed via the Sagnac effect,
\begin{equation}
\Delta\phi_{\rm Sagnac} = \frac{2\omega}{c^2}\int {\bf{\Omega}}\cdot d\mbox{\boldmath${\cal A}$}\;\;,
\end{equation}
where $\mbox{\boldmath${\cal A}$}$ refers to the area of the interferometer.  The measurement of this effect has led to many advances in basic and
applied physics [4].  In (21), $\omega$ is the frequency of electromagnetic waves or the de Broglie frequency of matter waves, i.e. $\omega = mc^2/\hbar$ to
lowest order.  To illustrate this point, it is interesting to derive (21) using a simple analogy with the Aharonov-Bohm effect [7].  The physical basis for
this analogy is the Larmor theorem, which for particles with the same charge-to-mass ratio $q/m$ establishes a local connection between phenomena in an
inertial frame but in the presence of a magnetic field ${\bf B}$ and those in a frame rotating at the Larmor frequency ${\bf{\Omega}} = q{\bf B}/(2mc)$. 
For instance, the canonical momentum of a particle of mass $m$ in a constant magnetic field ${\bf B}$ is ${\bf P} = m{\bf v} + q{\bf A}/c$, where $2{\bf A}
= {\bf B}\times {\bf r}$; similarly, the canonical momentum of a particle in a uniformly rotating frame is ${\bf P} = m{\bf v}^{\prime} +
m{\bf{\Omega}}\times {\bf r}$.  Therefore, the Sagnac phase shift can be obtained from the Aharonov-Bohm phase shift $q\oint{\bf A}\cdot d{\bf r}/(\hbar c)$
by replacing $q{\bf B}/(2c)$ with $m{\bf{\Omega}}$, i.e.
$\hbar\Delta\phi = m\oint({\bf{\Omega}}\times{\bf r})\cdot d{\bf r} = 2m{\bf{\Omega}}\cdot\mbox{\boldmath${\cal A}$}$.  

The Sagnac effect for matter waves was first directly measured by Werner {\it et al.} [8] via neutron interferometry.  Subsequently, it has been measured
for electrons [9] and atoms [10].  It is interesting to consider the theoretical extension of this effect to the gravitomagnetic field of rotating sources
using the gravitational Larmor theorem [11]. 

Gravitoelectromagnetism (``GEM'') usually refers to the various useful ways in which Einstein's general relativity can be described in terms of an analogy
with Maxwell's electrodynamics.  For our purposes, this approach is particularly helpful in the linear approximation of general relativity that is
applicable in the exterior of slowly moving astronomical bodies.  The classical tests of general relativity deal with certain post-Newtonian corrections to
the ``Newtonian'' gravitoelectric field that involves the mass-energy density.  The gravitomagnetic field involves the mass-energy current instead.  

The motion of particles in the gravitational field of a rotating mass was first studied by de Sitter [12] and later in a more general form by Thirring and
Lense [13]; however, GEM has a long history that dates back to the last decades of the nineteenth century [14].  Ciufolini [15] has described direct
evidence for the Lense-Thirring effect involving two $LAGEOS$ satellites in orbit about the Earth.  A direct measurement of the gravitomagnetic field of the
Earth is expected in the next few years from the GP-B, which involves superconducting gyroscopes in a polar orbit about the Earth [16].  The gravitomagnetic
precession frequency of an ideal gyroscope at rest outside a rotating mass is given by 
\begin{equation} {\bf{\Omega}}_P = \frac{GJ}{c^2r^3}\left[3({\bf{\hat J}}\cdot{\bf{\hat r}}){\bf{\hat r}} - {\bf{\hat J}}\right]\;\;,
\end{equation}
where $J$ is the angular momentum of the body and ${\bf r}$ is the position vector of the pointlike ideal gyroscope.  Let us note that the same
precession is locally obtained in Minkowski spacetime but in a frame rotating with a Larmor frequency of
$-{\bf{\Omega}}_{P}$.  This example illustrates the gravitational Larmor theorem. 

In close formal analogy with the classical Larmor theorem, a gravitomagnetic field can be locally replaced by a frame of reference in Minkowski spacetime
rotating with frequency ${\bf{\Omega}} = -{\bf{\Omega}}_P$.  It therefore follows from (21) with
${\bf{\Omega}}\rightarrow -{\bf{\Omega}}_P$ that 
\begin{equation}
\Delta\phi_{GM} = -\frac{2G\omega}{c^4}\oint\frac{({\bf J}\times{\bf r})\cdot d{\bf r}}{r^3}\;\;,
\end{equation}
where we have used the fact that $c^2{\bf{\Omega}}_{P} = {\bf{\nabla}}\times (G{\bf J}\times{\bf r}/r^3)$.  To examine the observational
possibilities for the gravitomagnetic effect (23), it is useful to begin by mentioning the corresponding gravitoelectric effects.  The first
gravitationally-induced quantum interference effect was observed in a neutron interferometer by Colella, Overhauser and Werner (``COW'') in 1975.  The COW
effect is given by 
\begin{equation}
\Delta\phi_{\rm COW} = \frac{g{\cal A}\omega}{c^2 v}\:{\rm sin}\:\alpha\;\;,
\end{equation}
where $g$ is the acceleration of gravity, ${\cal A}$ is the area of the interferometer, $\alpha$ is the angle between the plane of the
interferometer and the horizontal plane, $\omega$ is the de Broglie
frequency of the neutron and $v$ is its speed [17].  A similar formula can be derived for
light using classical general relativity, i.e. $\Delta\phi_{ph}=2g{\cal A}\omega\:{\rm sin}\:\alpha/c^3$; in fact, an experiment using fiber optics has been
proposed to measure this effect [18].  Similarly, for the case of electromagnetic radiation a space experiment [19] has been proposed to measure the
gravitomagnetic phase shift (23).  On the other hand, for matter waves the gravitomagnetic contribution --- due to the rotation of the Earth --- to the
interference phase shift (23) is smaller than the gravitoelectric phase shift (24) by
$\sim(v/c)(v_E/c)$, where $v_E$ is the equatorial speed of the rotating Earth.  It would be rather difficult in practice to separate the gravitomagnetic
effect from the dominant Sagnac effect (21) on the rotating Earth.  Nevertheless, it may be possible to measure the quantum gravitomagnetic effect (23) for
atoms using atomic interferometry [20]. 

Let us now turn our attention to the possibility of measuring the effects of intrinsic spin.  To this end, let us rewrite the Sagnac phase shift (21) in the
form
\begin{equation} 
\hbar\;\Delta\phi_{\rm Sagnac} = {\displaystyle
\sum_a}({\bf{\Omega}}\cdot{\bf L}_a)\Delta t_a \;, \end{equation}
where the index $a$ ranges over the various segments of the
interferometer and $\Delta t_a$ is a directional quantity defined by $\Delta{\bf r}_a = {\bf v}_a \; \Delta t_a$.  Since
$\sum_a\Delta{\bf r}_a = 0$, $\Delta t_a$ is positive (negative) along the counterclockwise (clockwise) beam.  Replacing
${\bf L}$ with the total angular momentum ${\bf L}+{\bf S}$ will in general result in an additional phase shift in the rotating frame due to the
spin-rotation coupling given by 
\begin{equation}
\Delta\phi_{SR} = \frac{1}{\hbar}\sum_a({\bf{\Omega}}\cdot{\bf S}_a)\Delta t_a\;.
\end{equation}
By using devices that would reverse the direction of spin on a segment of the interferometer, one could in principle measure the spin-rotation
phase shift (26).  For instance, if the two segments from the source to the detector are semicircular arcs of radius $r$, then for a particle of spin ${\bf
S} = \hbar\mbox{\boldmath$\sigma$}$ and speed $v$ we find $\Delta\phi_{SR} = 2\pi r\sigma\Omega/v$, assuming that on the counterclockwise (clockwise) arc
the spin is oriented parallel (antiparallel) to the axis of rotation of the interferometer.  Of course, the spin direction must be reversed again in one arc
just before the beams recombine in order that interference can take place.  It follows that the spin-rotation phase shift is proportional to the
circumference of the interferometer in contrast to the Sagnac phase shift that is proportional to the area of the interferometer.  For the sake of
comparison, we note that $\Delta\phi_{SR}/\Delta\phi_{\rm Sagnac} = \sigma\lambda/(2\pi r)$, so that
the spin-rotation effect is generally smaller than the Sagnac effect by the ratio of the (de Broglie) wavelength of the particle to the circumference of the
interferometer [21].  

Let us mention that in the gravitational field of a rotating mass, such as in a laboratory on the Earth, we expect a gravitomagnetic component of the
spin-rotation coupling as well, since the gravitational Larmor theorem implies that the effective rotation frequency is ${\bf{\Omega}} - {\bf{\Omega}}_P$,
where ${\bf{\Omega}}$ is the frequency of rotation of the Earth.
Therefore, the net result is 
\begin{equation}
\Delta\phi_{SR} =
\hbar^{-1}{\displaystyle \sum_a}({\bf{\Omega}} -
{\bf{\Omega}}_P)\cdot{\bf S}_a\;\Delta t_a\;.
\end{equation}
It is hoped that this effect could become measurable via 
optical or matter wave interferometry in the future.  

An interesting consequence of the spin-rotation coupling is a certain frequency-shift phenomenon.  Imagine a device and an incident particle such that in
passing through the device the particle's spin direction is reversed.  Consider such a device rotating with frequency ${\bf{\Omega}}$ in an inertial frame. 
Let $E_i$ be the initial energy of the incident particle and $E_f$ the final energy of the particle after passing through the rotating device.   For a
particle propagating parallel to the rotation axis of the device, the energy as measured by an observer at rest with the device is $E^{\prime}=E_i - {\bf
S}\cdot{\bf{\Omega}}$ as well as
$E^{\prime} = E_f + {\bf S}\cdot{\bf{\Omega}}$, so that $E_f - E_i = -2 \;{\bf S}\cdot{\bf{\Omega}}$.  For electromagnetic waves, the corresponding
frequency shift was apparently first discovered in the microwave regime [22].  Subsequently, the frequency shift has been studied in the optical regime by a
number of authors [23,24].  The general phenomenon of spin-rotation coupling for photons has been the subject of recent investigations [25].  The
spin-rotation energy shift would give rise to an observable beat phenomenon in a proposed interferometry experiment [5].

The consequences of spin-rotation coupling for atomic physics have been studied by Silverman [26]. Cai and Papini [27] have discussed the astrophysical
implications of this coupling for massive neutrinos.  Moreover, certain depolarization phenomena in circular accelerators [28] have been interpreted by
Papini {\it et al.} [29] in terms of the spin-rotation coupling.

For most laboratory applications, the spin-rotation interaction Hamiltonian involving intrinsic spin would be of the form $H_{int} = - {\bf
S}\cdot{\bf{\Omega}} + {\bf S}\cdot{\bf{\Omega}}_P$, where $\hbar\Omega\sim 10^{-19}$eV and
$\hbar\Omega_P\sim10^{-29}$ eV for a laboratory on the Earth.  Let us consider, for example, an experiment that involves the difference in the energy
$E$ of a particle of spin $S=\hbar\sigma$ polarized vertically up $(E_+)$ and down $(E_-)$ near the surface of an astronomical body.  Then, the interaction
under consideration implies that 
\begin{equation} E_+ - E_- = -2\sigma\;(\hbar\Omega_0)\;{\rm sin}\:\theta\;\;,
\end{equation}
where $\Omega_0 = \Omega - 2GJ/(c^2R^3), R$ is the radius of the body and
$\theta$ is the geographic latitude.  The $\Omega_0$ in (28) thus consists of two parts:  the rotation frequency of the body and a smaller gravitomagnetic
component.  There is observational support for the first term from high-precision experiments involving the coupling of the nuclear spin of Mercury to the
rotation of the Earth [30].  In connection with the second term, let us note that for Jupiter $\hbar(\Omega - \Omega_0)\simeq 10^{-27}$ eV, while for a
neutron star $\hbar(\Omega -
\Omega_0)\sim 10^{-14}$ eV. It is likely that with further improvements in magnetometer design, the spin-gravitomagnetic coupling could become measurable
near the surface of Jupiter in the foreseeable future [31].  It is important to recognize that such a relativistic quantum gravitational effect, like all
other gravitational effects, is subject to the whole mass-energy content of the universe.  Our treatment here has therefore been based on certain
cosmological assumptions regarding the distribution of angular momentum in the universe; specifically, we have assumed that on the largest scales there is
no preferred sense of rotation.  On the other hand, the observation of the gravitomagnetic effects under consideration here would imply certain interesting
restrictions on the distribution of mass-current in the universe.  

\section{Spin-rotation coupling, the Dirac equation and the Equivalence Principle}

Turning now to the case of spin $\frac{1}{2}$ particles, the relativistically correct way to treat this is to write the Dirac equation in a rotating frame of
reference.  Although the prescription for doing this is not perhaps widely known, it is perfectly well defined and is described in various places in the
literature [32-35].  The general form of the Dirac equation (in a coordinate basis) is
\begin{equation} 
i\hbar\gamma^{\lambda} \{\partial_{\lambda} + \frac{1}{8} [\gamma^{\mu} , \gamma^{\nu}]\; g _{\mu\rho}\; \Gamma^{\rho} _{\nu\lambda}\} \Psi =
m\Psi.
\end{equation}
Hehl and Ni [36] have calculated the resulting equation in an arbitrary noninertial frame --- one subject both to rotation and to acceleration. 
Expressing the equation in the form $i\hbar\partial\Psi/\partial t = H\Psi$ they have found an expression for the Hamiltonian $H$ in the nonrelativistic
limit, by taking appropriate Foldy-Wouthuysen transformations.  In the special case of zero acceleration (in which, therefore, the connection coefficients $
\Gamma^{\rho} _{\nu\lambda}$ are calculated essentially from the metric tensor (16) above) their Hamiltonian reduces to
\begin{equation} H = \beta  m + \beta \frac{p^2}{2m} - \omega. (\bf L+S).
\end{equation}
The first two terms, of course, represent rest mass and kinetic energy.  The next term, $\omega.\bf{L},$ represents a coupling of rotations with
orbital angular momentum and appears in the work of Page [37] and Werner {\it et al.} [8].  The final term, $\omega.{\bf S}$, is the spin-rotation coupling
whose existence was first mooted by Mashhoon [21]; see also [38].  An experiment to test for the existence of this term  has recently been proposed by
Mashhoon {\it et al.} [5].

Since the anticipated spin-rotation coupling term was found, as noted above, in a {\it non-relativistic approximation}, the question arises whether a
similar term appears in the general case.  This question is nontrivial
since the identification of the Pauli matrices ($\otimes 1$, to give
$4 \times 4$ matrices)
with the spin operators for spin $\frac {1}{2}$ particles only holds in the nonrelativistic limit.  A relativistic spin operator can, however, be defined,
and is in fact [39] closely related to the Foldy-Wouthuysen ``mean spin" operator.  It then follows easily [40] that the coupling of spin with rotations
holds also at the relativistic level.  (It should be noted, however, that this only holds in $\it Minkowski\;  space$.  Spin is defined, following Wigner
[41], as the ``little group'' of the inhomogeneous Lorentz group; that is, the group which leaves a given 4-momentum invariant.  The inhomogeneous Lorentz
group, however, is the isometry group of Minkowski space; in a general Riemannian space there is no isometry group, so spin is undefined.)

A further question which presents itself is concerned with the
\emph{Equivalence Principle}.  In Einstein's traditional argument,
(noninertial) 
accelerations are equivalent to a gravitational field, so we might expect, assuming that the Equivalence Principle holds down to the quantum level, the same
physical effects to emerge from the behavior of Dirac particles in an accelerating reference frame and in a Schwarzschild spacetime --- that is, after due
attention has been paid to the fact that the Equivalence Principle (EP) is a $\it local$ one, in which tidal forces are neglected.  If we now envisage  a
generalization of the EP to include (noninertial) $\it rotations$, we should then expect similar physical effects from a Dirac particle in a general
noninertial frame with one in a Kerr spacetime.  Some investigations have recently been made into this interesting area [42-44].


\begin{thebibliography}{99}
\bibitem{[1]}  A. Einstein, {\it The Meaning of Relativity} (Princeton University Press, Princeton, 1955).

\bibitem{[2]} B. Mashhoon, Found. Phys. (Wheeler Festschrift) {\bf 16}, 619 (1986); Gen. Rel. Grav. {\bf 31}, 681 (1999).

\bibitem{[3]} B. Mashhoon, Phys. Rev. A {\bf 47}, 4498 (1993); U. Muench, F. W. Hehl and B. Mashhoon, Phys. Lett. A {\bf 271}, 8 (2000); B. Mashhoon,
``Relativity and Nonlocality,'' gr-qc/0011013.

\bibitem{[4]} R. Anderson, H. R. Bilger and G. E. Stedman, Am. J. Phys. {\bf 62}, 975 (1994); G. E. Stedman, Contemp. Phys. {\bf 26}, 311 (1985); G. E.
Stedman, Rep. Prog. Phys. {\bf 60}, 615 (1997).

\bibitem{[5]} B. Mashhoon, R. Neutze, M. Hannam and G. E. Stedman, Phys. Lett. A {\bf 249}, 161 (1998).

\bibitem{[6]} B. Mashhoon, Phys. Lett. A {\bf 139}, 103 (1989); Class. Quantum Grav. {\bf 17}, 2399 (2000).

\bibitem{[7]} J. J. Sakurai, Phys. Rev. D {\bf 21}, 2993 (1980).

\bibitem{[8]} S. A. Werner, J.-L. Staudenmann and R. Colella, Phys. Rev. Lett. {\bf 42}, 1103 (1979); H. Rauch and S. A. Werner, {\it Neutron Interferometry}
(Clarendon Press, Oxford, 2000).

\bibitem{[9]} F. Hasselbach and M. Nicklaus, Phys. Rev. A {\bf 48}, 143 (1993).

\bibitem{[10]} T. L. Gustavson, A. Landragin and M. A. Kasevich, Class. Quantum Grav. {\bf 17}, 2385 (2000).

\bibitem{[11]} B. Mashhoon, Phys. Lett. A {\bf 173}, 347 (1993); J. M. Cohen and B. Mashhoon, Phys. Lett. A {\bf 181}, 353 (1993).

\bibitem{[12]} W. de Sitter, Mon. Not. Roy. Astron. Soc. {\bf 76}, 699 (1916).

\bibitem{[13]} H. Thirring, Phys. Z. {\bf 19}, 33 (1918); {\bf 22}, 29 (1921); J. Lense and H. Thirring, Phys. Z. {\bf 19}, 156 (1918); B. Mashhoon, F. W. Hehl
and D. S. Theiss, Gen. Rel. Grav. {\bf 16}, 711 (1984). 

\bibitem{[14]} B. Mashhoon, F. Gronwald and H. I. M. Lichtenegger, in {\it Testing Relativistic Gravity in Space}, edited by C. L\"{a}mmerzahl, C. W. F.
Everitt and F. W. Hehl (Springer, Berlin, 2001). 

\bibitem{[15]} I. Ciufolini, Class. Quantum Grav. {\bf 17}, 2369 (2000).

\bibitem{[16]} C. W. F. Everitt {\it et al.}, in {\it Testing Relativistic Gravity in Space}, edited by C. L\"{a}mmerzahl, C. W. F. Everitt and F. H. Hehl
(Springer, Berlin, 2001).

\bibitem{[17]} R. Colella, A. W. Overhauser and S. A. Werner, Phys. Rev. Lett. {\bf 34}, 1472 (1975); H. Rauch and S. A. Werner, {\it Neutron Interferometry}
(Clarendon Press, Oxford, 2000); K. Varj\'{u} and L. H. Ryder, Phys. Lett. A {\bf 250}, 263 (1998); Phys. Rev. D {\bf 62}, 024016 (2000);  Am. J. Phys. {\bf
68}, 404 (2000).

\bibitem{[18]} K. Tanaka, Phys. Rev. Lett. {\bf 51}, 378 (1983); J. M. Cohen and B. Mashhoon, Phys. Lett. A {\bf 181}, 353 (1993).

\bibitem{[19]} R. W. Davies, in: {\it Experimental Gravitation}, edited by B. Bertotti (Academic Press, New York, 1974), p. 405; R. W. Davies and H. Lass, JPL
Technical Memorandum no. 58 (1970).

\bibitem{[20]} C. J. Bord\'{e}, Phys. Lett. A {\bf 140}, 10 (1989); M. Kasevich and S. Chu, Appl. Phys. B {\bf 54}, 321 (1992); J. Audretsch and C.
L\"{a}mmerzahl, Appl. Phys. B {\bf 54}, 351 (1992).

\bibitem{[21]} B. Mashhoon, Phys. Rev. Lett. {\bf 61}, 2639 (1988); {\bf 68}, 3812 (1992).

\bibitem{[22]} P. J. Allen, Am. J. Phys. {\bf 34}, 1185 (1966).

\bibitem{[23]} P. Crane, Appl. Opt. {\bf 8}, 538 (1969); B. A. Garetz and S. Arnold, Opt. Commun. {\bf 31}, 1 (1979); B. A. Garetz, J. Opt. Soc. Am. {\bf 71},
609 (1981). 

\bibitem{[24]} M. P. Kothiyal and C. Delisle, Opt. Lett. {\bf 9}, 319 (1984); J. P. Hugnard and J. P. Herriau, Appl. Opt. {\bf 24}, 4285 (1985);  R. Simon, H.
J. Kimble and E. C. G. Sudarshan, Phys. Rev. Lett. {\bf 61}, 19 (1988); F. Bretenaker and A. le Floch, Phys. Rev. Lett. {\bf 65}, 2316 (1990); V. Begini
{\it et al.}, Eur. J. Phys. {\bf 15}, 71 (1994); R. Bhandari, Phys. Rep. {\bf 281}, 1 (1997).

\bibitem{[25]} J. Courtial {\it et al.}, Phys. Rev. Lett. {\bf 80}, 3217 (1998);  Phys. Rev. Lett. {\bf 81}, 4828 (1998).

\bibitem{[26]} M. P. Silverman, Phys. Lett. A {\bf 152}, 133 (1991); Nuovo Cimento D {\bf 14}, 857 (1992).

\bibitem{[27]} Y. Q. Cai and G. Papini, Phys. Rev. Lett. {\bf 66}, 1259 (1991); {\bf 68}, 3811 (1992).

\bibitem{[28]} J. S. Bell and J. Leinaas, Nucl. Phys. B {\bf 212}, 131 (1983); {\bf 284}, 488 (1987).

\bibitem{[29]} Y. Q. Cai , D. G. Lloyd and G. Papini, Phys. Lett. A {\bf 178}, 225 (1993); D. Singh and G. Papini, Nuovo Cimento B {\bf 115}, 223 (2000).

\bibitem{[30]} B. J. Venema {\it et al.}, Phys. Rev. Lett. {\bf 68}, 135 (1992); D. J. Wineland {\it et al.}, Phys. Rev. Lett. {\bf 67}, 1735 (1991); B.
Mashhoon, Phys. Lett. A {\bf 198}, 9 (1995).

\bibitem{[31]} B. Mashhoon, Class. Quantum Grav. {\bf 17}, 2399 (2000).

\bibitem{[32]} D. R. Brill and J. A. Wheeler,  Rev. Mod. Phys. {\bf 29}, 465 (1957).

\bibitem{[33]} D. R. Brill and J. M. Cohen,  J. Math. Phys. {\bf 7}, 238 (1966).

\bibitem{[34]} R. U. Sexl and H. K. Urbantke, {\it Gravitation und Kosmologie} (Bibliographisches Institut, Mannheim, 1983), p. 311.
 
\bibitem{[35]} T. Frenkel, {\it The Geometry of Physics} (Cambridge University Press, Cambridge, 1997), p. 491.

\bibitem{[36]} F. W. Hehl and W.-T. Ni,  Phys. Rev. D {\bf 42}, 2045 (1990).

\bibitem{[37]} L. A. Page,  Phys. Rev. Lett. {\bf 35}, 543 (1975).

\bibitem{[38]} I. Damiao Soares and J. Tiomno,  Phys. Rev. D {\bf 54}, 2808 (1996).

\bibitem{[39]} L. H. Ryder, {\it these Proceedings}.

\bibitem{[40]} L. Ryder,  J. Phys. A {\bf 31}, 2465 (1998).

\bibitem{[41]} E. P. Wigner,  Ann. Math. {\bf 40}, 149 (1939).

\bibitem{[42]} K. Varj\'{u} and L. H. Ryder,  Phys. Rev. D {\bf 62}, 024016 (2000).

\bibitem{[43]} G. Z. Adunas, E. Rodriguez-Milla and D. V. Ahluwalia, ``Probing Quantum Violations of the Equivalence Principle", Gravity Research Foundation
Essay (2000); Phys. Lett. B {\bf 485}, 215 (2000).

\bibitem{[44]} Yu. N. Obukhov, Phys. Rev. Lett. {\bf 86}, 192 (2001). 
 \end{thebibliography}
\end{document}